\documentstyle[aps,epsfig,multicol]{revtex}
\begin{document}
\draft
\title{Quantum phase transitions in superconducting arrays under external magnetic fields}

\author{Beom Jun Kim$^{1,*}$, Gun Sang Jeon$^1$, 
M.-S. Choi$^{2}$, and M.Y. Choi$^1$}
\address{
    $^1$Department of Physics and Center for Theoretical Physics \\
     Seoul National University,
     Seoul 151-742, Korea}
\address{
    $^2$Department of Physics and Pohang Superconductivity Center \\
     Pohang University of Science and Technology,
     Pohang 790-784, Korea}

\preprint{SNUTP-051}
\maketitle
\thispagestyle{empty}

\begin{abstract}
We study the zero-temperature phase transitions of two-dimensional
superconducting arrays with both
the self- and the junction capacitances
in the presence of external magnetic fields. 
We consider two kinds of excitations from the Mott insulating phase: 
charge-dipole excitations and single-charge excitations, 
and apply the second-order perturbation theory to find their
energies. 
The resulting phase boundaries are found to depend strongly 
on the magnetic frustration, which measures 
the commensurate-incommensurate effects. 
Comparison of the obtained values with those in recent experiment suggests
the possibility that the superconductor-insulator transition observed in 
experiment may not be of the Berezinskii-Kosterlitz-Thouless type.
The system is also transformed to a classical three-dimensional
$XY$ model with the magnetic field in the time-direction;
this allows the analogy to bulk superconductors,
revealing the nature of the phase transitions. 
\end{abstract}


\begin{multicols}{2}

\section{Introduction}
Two-dimensional (2D) superconducting arrays with charging effects have 
drawn much 
interest because of their interesting phase transitions~\cite{proceeding}
and relations to other systems such as the Bose-Hubbard model and
the quantum $XXZ$ spin model~\cite{otterlo1}.
In those arrays where both the self-capacitance $C_0$ 
and the junction capacitance $C_1$ are present, charging energies
due to both capacitances need to
be considered simultaneously since the nature of the phase transition
in general depends on them. 
For example, at zero temperature the system with only self-capacitance can be
mapped into a classical three-dimensional (3D) $XY$ model 
with the ratio $E_0/E_{\rm J}$ of the
charging energy $E_0 \equiv e^2/2C_0$ 
to the Josephson coupling strength $E_{\rm J}$ taking
the role of the temperature.
In the opposite limit, on the other hand, 
it is well known that there exists an interesting duality between charges 
and vortices~\cite{Duality,fazio}, and the system with only junction capacitance
at zero temperature undergoes a
charge-unbinding Berezinskii-Kosterlitz-Thouless (BKT) 
transition~\cite{BKT} from the insulating phase to the 
superconducting one as the ratio $E_{\rm J}/E_1$ with $E_1 \equiv e^2/2C_1$
is increased.
The critical value $(E_{\rm J}/E_1)_{\rm c}$, beyond which the array is 
superconducting,  has been found $0.6$ 
in experiment~\cite{zant},  $0.23$ in the 
duality argument~\cite{fazio} and in the variational method~\cite{cont}, 
$0.51$ in perturbation expansion~\cite{b0},
and $0.36$ in quantum Monte Carlo simulations~\cite{rojas}. 
As $C_0$ is increased from zero in this system,
the interactions between charges are screened, with the screening length
given by $\Lambda \equiv \sqrt{C_1/C_0}$~\cite{fazio}, and 
the nature of the phase transition is expected to alter.
Recently, the zero-temperature phase diagrams have been studied by means of 
the perturbation expansion~\cite{b0}, which suggests that 
the phase transition from the Mott insulating phase to the superconducting
phase is governed by the single-charge (SC) excitations or by the 
charge-dipole (CD) excitations, depending on the ratio $C_1/C_0$ as well as
on the charge frustration.

In this paper, we study the quantum phase transitions in 2D
superconducting arrays, focusing on the effects of external magnetic fields
as well as the competition between self- and junction capacitances.
The effects of external magnetic fields on phase transitions have been 
studied both
in the classical arrays without charging energy~\cite{proceeding} and in
the quantum arrays~\cite{rojas,kampf,granato,bruder}.
However, most existing analytical works on the latter have employed 
mean-field-like approximations, which are not reliable in two 
dimensions. 
We thus adopt the perturbative expansion instead, and consider the 
SC and the CD excitations 
to the second order in $E_{\rm J}$.
The obtained phase boundaries between the Mott insulating phase 
and the superconducting one exhibit strong commensurability effects 
due to the magnetic frustration.
The results are compared with experiments, suggesting the possibility
that the experimentally observed transition may not be of the 
charge-unbinding BKT type.
It is also shown that the dual transformation maps the
system with both magnetic frustration and general capacitance
onto a classical 3D $XY$ model under the magnetic field 
in the time-direction.  
This transformation allows us to discuss the nature of the phase transitions,
by analogy with bulk superconductors under magnetic fields.

There are four sections in this paper: Section~II is devoted to the
perturbative approach, from which the zero-temperature diagrams are 
obtained. 
We compare the phase diagrams with those obtained from the mean-field
approach and those observed in experiments.
In Sec.~III the system is transformed
into a classical 3D $XY$ model under the magnetic field in the 
(imaginary) time direction.
The nature of the phase transitions is discussed
by analogy with bulk superconductors under magnetic fields.
Finally, Sec.~IV summarizes the results and presents some discussions.

\section{Perturbative Approach}

We begin with the Hamiltonian describing the superconducting array
with magnetic frustration:
\begin{eqnarray} \label{eqH}
  H &=& 4E_0\sum_{i,j} q_i {\widetilde C}_{ij}^{-1}q_j
      -E_{\rm J}\sum_{\langle i,j \rangle}\cos (\phi_i - \phi_j -A_{ij})\\
\nonumber &\equiv& H_0 + V ,
\end{eqnarray}
where the charge $q_j$ at site $j$ represented in units of 
$2e$ ($e>0$) and the phase $\phi_k$ of the superconducting order parameter
at site $k$ satisfy the commutation relation $[\phi_k,q_j] = i\delta_{jk}$. 
Whereas the summation in the second term runs over all nearest neighboring
pairs, the charges interact via the inverse of the dimensionless capacitance matrix 
${\widetilde C}_{ij}$ defined by 
\end{multicols}
\noindent\rule{0.5\textwidth}{0.1ex}\rule{0.1ex}{2ex}\hfill
\begin{equation}
{\widetilde C}_{ij} \equiv (1 + 4C_1/C_0)\delta_{i,j} - (C_1/C_0)
( \delta_{i,j+\hat{\bf x}} +\delta_{i,j-\hat{\bf x}} + \delta_{i,j+\hat{\bf y}}
+ \delta_{i,j-\hat{\bf y}} ) 
\end{equation}
\hfill\raisebox{-1.9ex}{\rule{0.1ex}{2ex}}\rule{0.5\textwidth}{0.1ex}
\begin{multicols}{2}
\noindent
with the self-capacitance $C_0$ and the junction capacitance $C_1$.
The important external parameters $E_0 \equiv e^2/C_0$ and  $E_{\rm J}$ 
measure the self-charging energy and the Josephson coupling energy, respectively. 
The magnetic bond angle $A_{ij}$ between the two sites $i$ and $j$ is given by
the line integral of the magnetic vector potential $\bf {A}$:
$A_{ij} \equiv (2\pi / \Phi_0)\int_i^j {\bf A} \cdot d {\bf l}$ with 
the magnetic flux quantum $\Phi_0 \equiv hc/2e$. 

When $E_{\rm J}=0$, the system described by the 
unperturbed Hamiltonian $H_0$ has the Mott insulating phase as the ground state 
with the charge configuration $q_i = 0$ at any site.
In the opposite limit of $E_0 = 0$, on the other hand, the system 
is described by the 
2D classical $XY$ model, displaying superconductivity at zero temperature.
It is thus expected that there exists a critical value of $E_{\rm J}/E_0$ 
beyond which the superconducting phase becomes energetically favorable.
The critical value may be determined by comparing the energy of the Mott
insulating phase with that of the superconducting phase.
The energy $E_{\rm M}$ of the Mott insulating phase is easily computed
up to the second order in $E_{\rm J}/E_0$:
\begin{equation} \label{eq_EM}
E_{\rm M} = - \frac{E_{\rm J}^2 N}
   {8E_0 ({\widetilde C}^{-1}_{00} - {\widetilde C}^{-1}_{\hat{\bf x}, 0})}
\end{equation}
for the $L\times L$ square array $(N \equiv L^2)$
(see Ref.~\onlinecite{b0} for the detailed calculation).
It is of interest to note here that $E_{\rm M}$ does not depend on the
external magnetic field.

\begin{figure}
\centerline{\epsfig{file=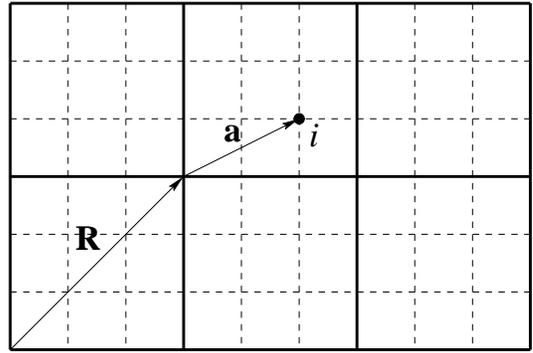,width=7.0cm}}
\vskip 1cm
\caption{Superconducting array of size $9 \times 6$ with 
the magnetic frustration $f=1/3$. 
The $3\times 3$ superlattices are indicated by thick solid lines. 
The position of the $i$th lattice site is represented  by 
${\bf R} + {\bf a}$, where ${\bf R}$ and ${\bf a}$ denote the position 
of the superlattice and the relative position of the site inside the 
superlattice, respectively.}
\label{fig_vector}
\end{figure}

Since the lowest excitation that can lead to the change of the ground state
from the Mott insulating into the superconducting phase is presumably 
point-like, and we consider two kinds of excitations:
the SC and the CD excitations.
We first consider the SC-type excited state
and write its energy up to the second order in $E_{\rm J}/E_0$:
\begin{equation}
E_{\rm SC} = E_{\rm SC}^{(0)} + E_{\rm SC}^{(1)} + E_{\rm SC}^{(2)}.
\end{equation}
In the SC-type excited state a single positive charge is located only 
at one site, and the zeroth-order energy is given by
\begin{equation}
E_{\rm SC}^{(0)} = 4E_0 \sum_{i,j} q_i \widetilde{C}^{-1}_{ij}q_j 
 = 4E_0\widetilde{C}^{-1}_{00}.
\end{equation}
Since the single charge can be located at any site without any energy
difference, we need to make use of the degenerate perturbation theory.
In the first order it is thus necessary to diagonalize the matrix
\begin{equation} \label{eq_Vij}
\langle i|V|j\rangle = - \langle i | E_{\rm J} \sum_{\langle k,l\rangle} 
\cos(\phi_k {-} \phi_l {-} A_{kl})|j\rangle 
\equiv -\frac{E_{\rm J}}{2} P_{ij} ,
\end{equation}
where $|i\rangle$ denotes the charge eigenstate with the single charge at site $i$
and the matrix element $P_{ij}$ is defined by
\end{multicols}
\noindent\rule{0.5\textwidth}{0.1ex}\rule{0.1ex}{2ex}\hfill
\begin{displaymath}
P_{ij} \equiv \left\{ \begin{array}{cl}
            \exp(-iA_{ij}) & \ \ \ 
          \mbox{for nearest neighboring sites $i$ and $j$, } \\
            0 & \ \ \ \mbox{otherwise.}
         \end{array} \right.
\end{displaymath}
\hfill\raisebox{-1.9ex}{\rule{0.1ex}{2ex}}\rule{0.5\textwidth}{0.1ex}
\begin{multicols}{2}
\noindent
In the presence of the external magnetic field, the gauge-invariant magnetic 
frustration is defined by $ f \equiv \Phi /\Phi_0$ with $\Phi$ being the magnetic 
flux per plaquette.  If $f = p/q$ with $p$ and $q$ being relative primes, it 
is obvious that $P_{ij}$ is invariant under the magnetic translation 
of $q$ lattice sites.   
Noting this translational symmetry, we represent the position of site $i$
by the vector ${\bf R}{+}{\bf a}$, where ${\bf R}$ is the position vector of 
the $q\times q$ superlattice unit cell containing site $i$ and ${\bf a}$
denotes the relative position of site $i$ inside the superlattice (see Fig.~\ref{fig_vector}), and
write 
\begin{equation}
P_{ij} = P({\bf R},{\bf a}; {\bf R}^{\prime}, {\bf a}^{\prime}) 
 = P({\bf R}{-}{\bf R}^{\prime}, {\bf a}; {\bf 0}, {\bf a}^{\prime}).
\end{equation}
Through the Fourier transformation
\begin{eqnarray} \nonumber
{\bar v}^{({\bf p})} ({\bf a}) &\equiv& \frac{1}{\sqrt{M}} \sum_{\bf R} 
e^{i {\bf p}\cdot {\bf R}} \,v({\bf R}{+}{\bf a} ) , \label {eq_FT_v} \\
{\bar P}^{({\bf p})} ({\bf a} ; {\bf a}^{\prime}) &\equiv& \sum_{\bf R} 
e^{i {\bf p}\cdot {\bf R}} P({\bf R}, {\bf a} ; {\bf 0}, {\bf a}^{\prime}) ,
\end{eqnarray}
the matrix is block-diagonalized, resulting in 
the eigenvalue equations 
\begin{equation}
\sum_{{\bf a}^{\prime} } {\bar P}^{({\bf p})} ({\bf a}; {\bf a}^{\prime}) 
{\bar v}^{({\bf p})} ({\bf a}^{\prime})
  =  \lambda^{({\bf p})} {\bar v}^{({\bf p})} ({\bf a}),
\end{equation}
where $M \equiv N/q^2$ is the total number of superlattices, 
and $v({\bf R}{+}{\bf a}) \equiv v_i$ is the wavefunction for the 
eigenstate of $P_{ij}$.
The numerical diagonalization of $M$ Hermitian matrices 
${\bar P}^{({\bf p})} ({\bf a} ; {\bf a^\prime} )$, each of the size 
$q^2 \times q^2$, 
yields the largest eigenvalue $P_{\rm max}$, which in turn gives the
first-order energy
\begin{equation}
E_{\rm SC}^{(1)} = - E_{\rm J} P_{\rm max} / 2 .
\end{equation}

\begin{figure}
\centerline{\epsfig{file=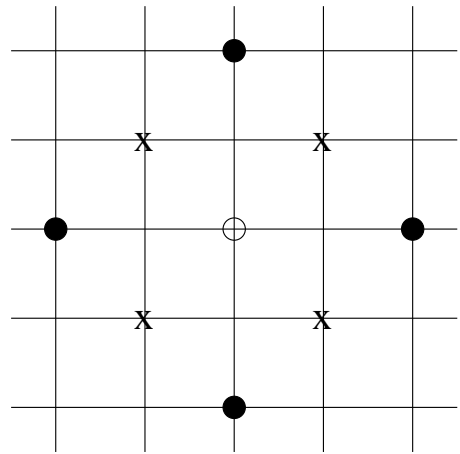,width=6.0cm}}
\vskip 1cm
\caption{Three cases in which the second-order matrix $\widetilde{Q}_{ij}$ 
has nonzero values. 
Only the single-charge states $|j\rangle$ with $j$ at positions 
marked by the empty circle,
the filled circles, and the `x' symbols, can be connected 
by two successive charge hoppings, to the state $|i\rangle$
with $i$ at the center ($\circ$).}
\label{fig_sc_schem}
\end{figure}

\begin{figure}
\centerline{\epsfig{file=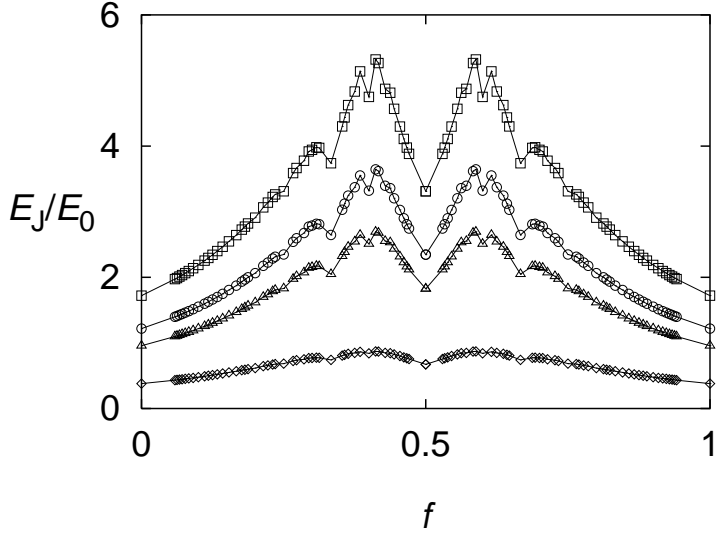,width=10.0cm}}
\vskip 1cm
\caption{Phase boundaries between the Mott insulating phase (below
each curve) and the superconducting phase (above), 
computed from the consideration of single-charge excitations.
Boundaries for various ratios of the junction capacitance $C_1$ to the
self-capacitance $C_0$ are shown: $C_1/C_0$ = 0.0001($\Box$),
0.1($\circ$), 0.2($\triangle$), and 1.0($\Diamond$). 
It is observed that the superconducting region expands as the junction 
capacitance is increased.}
\label{fig_sc}
\end{figure}

It is well known that the
largest eigenvalue $P_{\rm max}$ always shows up for ${\bf p} = 0$,
and further,
the eigenstates corresponding to the eigenvalue $P_{\rm max}$ are 
$q$-fold degenerate for $f=p/q$~\cite{mychoi85}.
Since the degeneracy is not completely removed in the first-order calculation, 
it is necessary to diagonalize the second-order matrix $Q$ with the element given by
\begin{equation} \label{eq_Qdd}
Q_{d d^\prime} = \sum_{ {\bf q}  \notin D} \frac{ \langle d | V | {\bf q}  \rangle 
\langle {\bf q}  | V | d^\prime \rangle} {E_{\rm SC}^{(0)} - E_{\bf q}^{(0)} } ,
\end{equation}
where the summation runs over all the charge eigenstates 
$| {\bf q} \rangle$ outside the space 
$D$ spanned by the SC states, and $| d \rangle $ represents the $d$th 
eigenstate corresponding to the eigenvalue $P_{\rm max}$: 
\begin{equation} \label{eq_d}
| d \rangle = \sum_i v_{d,i} | i \rangle = \sum_{ {\bf R}, {\bf a} } 
v_d({\bf R}{+}{\bf a}) | {\bf R}{+}{\bf a} \rangle .
\end{equation}
Here the wavefunction $v_d({\bf R}{+}{\bf a})$ of the $d$th degenerate state
is related to the component ${\bar v}_d({\bf a})$ of the normalized 
eigenvector of $\bar P$ via \begin{equation} \label{eq_vd}
v_d({\bf R}{+}{\bf a}) = \frac{1}{\sqrt{M}} {\bar v}_d({\bf a}) ,
\end{equation}
where the superscript in ${\bar v}_d^{{\bf p} = 0 }({\bf a})$ has been
omitted for simplicity [see Eq.~(\ref{eq_FT_v})].
Inserting Eqs.~(\ref{eq_d}) and (\ref{eq_vd}) into Eq.~(\ref{eq_Qdd}), we
obtain
\begin{equation} \label{eq_Qdd1}
Q_{dd^\prime} = \sum_{\bf R} \sum_{ {\bf a}, {\bf a^\prime} } {\bar v}_d^* ({\bf a})
{\bar v}_{d^\prime} ({\bf a^\prime} ) \widetilde{Q}_{ {\bf R} + {\bf a}, {\bf a^\prime}} 
\end{equation}
with 
\begin{displaymath} 
\widetilde{Q}_{{\bf R} + {\bf a}, {\bf a^\prime}} \equiv
 \sum_{ {\bf q} \notin D} \frac{ \langle i | V |  {\bf q} \rangle \langle  {\bf q} | V | j \rangle }
 { E_{\rm SC}^{(0)} - E_ {\bf q}^{(0)} }\equiv \widetilde{Q}_{ij}.
\end{displaymath}
It is obvious that $\widetilde{Q}_{ij}$ does not vanish 
only when $|i\rangle$ and
$|j\rangle$ are related by two successive charge hoppings as shown in 
Fig.~\ref{fig_sc_schem}. 
When $i=j$ (denoted by an empty circle in Fig.~\ref{fig_sc_schem}), 
it is easy to find that
\begin{equation} \label{eq_wideQ1}
\widetilde{Q}_{ii} = \frac{E_{\rm J}^2}{32 E_0} {\sum_{(m,n)}}^\prime
\frac{1}{ (\widetilde{C}^{-1}_{ni} - \widetilde{C}^{-1}_{mi}) 
 - (\widetilde{C}^{-1}_{00} - \widetilde{C}^{-1}_{\hat{\bf x}, 0})} , 
\end{equation}
where the summation ${\sum^\prime_{(m,n)}}$ runs over $m$ and its four nearest 
neighbors $n$ with the restriction of nonzero denominator.
When $j = i\pm \hat{\bf x} \pm \hat{\bf y}$ (denoted by `x' symbols in 
Fig.~\ref{fig_sc_schem}), there exist two intervening states $| {\bf q} \rangle$,
yielding
\begin{equation} \label{eq_wideQ2}
\widetilde{Q}_{i, j = i \pm \hat{\bf x} \pm \hat{\bf y} } = 
\frac{E_{\rm J}^2}{32 E_0} 
\frac{ e^{i ( A_{j,i\pm\hat{\bf x}} + A_{i\pm\hat{\bf x}, i} )} +
       e^{i ( A_{j,i\pm\hat{\bf y}} + A_{i\pm\hat{\bf y}, i} )}  }
     { (\widetilde{C}^{-1}_{\hat{\bf x}, 0} - \widetilde{C}^{-1}_{\hat{\bf x}+\hat{\bf y},0})
      - ( \widetilde{C}^{-1}_{00} - \widetilde{C}^{-1}_{\hat{\bf x}, 0}) } .
\end{equation}
On the other hand, when $j = i \pm 2 \hat{\bf x} ( 2 \hat{\bf y})$ as
represented by filled circles in Fig.~\ref{fig_sc_schem}, 
the matrix element is computed to be
\begin{equation} \label{eq_wideQ3}
\widetilde{Q}_{i, j = i \pm 2\hat{\bf x} ( 2\hat{\bf y} ) } =
\frac{E_{\rm J}^2}{32 E_0}
\frac{ e^{i ( A_{j,i\pm  \hat{\bf x} (\hat{\bf y}) } + 
                  A_{i\pm \hat{\bf x} (\hat{\bf y} ), i} ) } }
     { (\widetilde{C}^{-1}_{\hat{\bf x}, 0} - \widetilde{C}^{-1}_{2\hat{\bf x},0})
      - ( \widetilde{C}^{-1}_{00} - \widetilde{C}^{-1}_{\hat{\bf x}, 0}) } .
\end{equation}
Equations~(\ref{eq_wideQ1}) -- (\ref{eq_wideQ3}) together with the eigenvector 
$\bar v$ obtained in the first-order calculation yield the explicit form of 
the matrix element $Q_{dd^\prime}$ in Eq.~(\ref{eq_Qdd1}). 
The matrix $Q$ is then diagonalized to give 
the minimum eigenvalue $Q_{\rm min}$, which in turn leads to
the energy of the SC state:
\end{multicols}
\noindent\rule{0.5\textwidth}{0.1ex}\rule{0.1ex}{2ex}\hfill
\begin{equation}
E_{\rm SC} = E_{\rm SC}^{(0)} + E_{\rm SC}^{(1)} + E_{\rm SC}^{(2)} =
  4E_0\widetilde{C}^{-1}_{00} - E_{\rm J} P_{\rm max} / 2 + Q_{\rm min}.
\end{equation}
\hfill\raisebox{-1.9ex}{\rule{0.1ex}{2ex}}\rule{0.5\textwidth}{0.1ex}
\begin{multicols}{2}
\noindent
Comparing it with the energy of the Mott phase in Eq.~(\ref{eq_EM}), 
we find the phase boundary between the Mott insulating phase and 
the superconducting phase.
Figure~\ref{fig_sc} shows the obtained phase boundaries separating the
superconducting phase (above each curve) and the insulating one 
(below) on the plane of $E_{\rm J}/E_0$ and $f$, for various values of 
$C_1/C_0$. 
In obtaining the critical values $(E_{\rm J}/E_0)_{\rm c}$ numerically,
we have considered systems of sufficiently large sizes, so that
the critical values display convergence in at least three significant digits,
for given parameters $f$ and $C_1/C_0$.
Thus the system size has been increased up to $N=384$, depending on
the values of $f$ and $C_1/C_0$, and the convergence has been confirmed.

It is obvious that the superconducting region expands as the junction 
capacitance $C_1$ is increased, confirming the results 
in Refs.~\cite{cont,b0}.
In particular, the obtained phase diagrams are entirely similar to 
those obtained
for $C_0 \gg C_1$ in the mean-field approximation~\cite{bruder},
demonstrating significant commensurate-incommensurate effects due to the
magnetic frustration. 
Quantitatively, however, there exists discrepancy between the results of
the perturbation expansion and those from the mean-field approaches: 
The estimated critical values in the former are rather larger.
Furthermore, the perturbation expansion yields the ratio of the 
critical value for $f=1/2$ to that for
$f=0$ approximately given by 1.9 for $C_1/C_0 \lesssim
0.1$ (see Fig.~3); this is larger than the value $\sqrt{2}$ predicted in the
self-charging limit within the mean-field
approximation~\cite{granato,bruder}.
The increase of $C_1/C_0$ is found to reduce the ratio monotonically.
It is of interest to notice here that the first-order calculation reproduces
the mean-field value $\sqrt{2}$ regardless of $C_1/C_0$.

\begin{figure}
\centerline{\epsfig{file=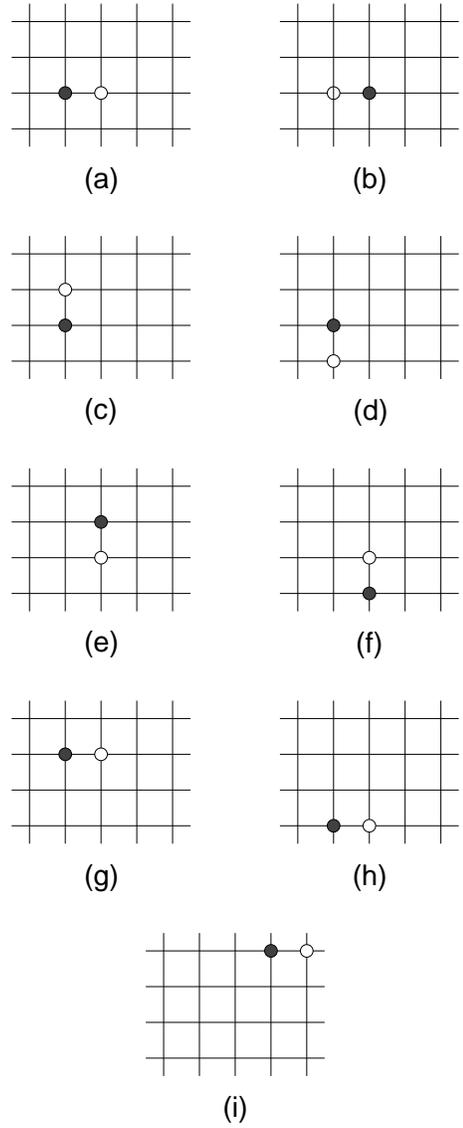,width=6.0cm}}
\vskip 1cm
\caption{The charge-dipole-type states $|k,l\rangle$ 
giving nonzero elements of the second-order matrix 
$ \langle i,j|M|k,l \rangle$
in case that $|i,j\rangle$ has the charge configuration 
of a positive charge at site $i$ and a negative charge at $j$ as shown in (a). 
The filled and the empty circles denote the positive and the negative charges,
respectively.
(i) shows an example of the states not included in (a)-(h). }
\label{fig_cd_schem}
\end{figure}

\begin{figure}
\centerline{\epsfig{file=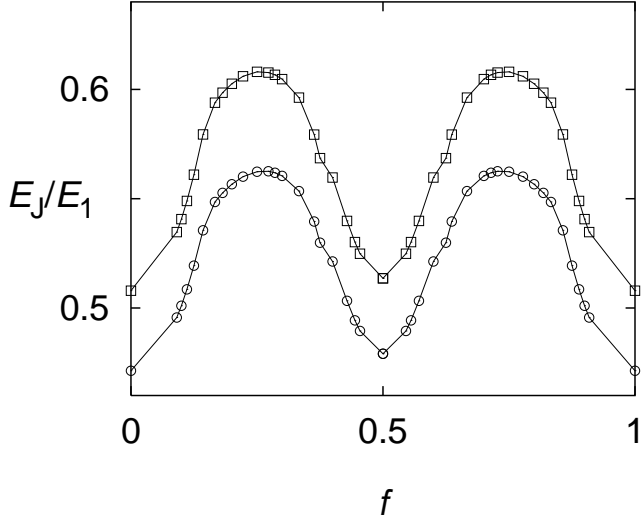,width=10.0cm}}
\vskip 1cm
\caption{Phase boundaries between the Mott insulating phase (below
each curve) and the superconducting phase (above)
of the charge-dipole type excitation.
The upper curve is for $C_1/C_0 = 10000$ while the lower one is 
for $C_1/C_0 = 10$.
}
\label{fig_cd}
\end{figure}

We now consider the CD-type excited state, where there exists a pair of 
positive and negative charges separated by the lattice spacing.
The energy of the CD state is written as
\begin{equation}
E_{\rm CD} = E_{\rm CD}^{(0)} +  E_{\rm CD}^{(1)} +  E_{\rm CD}^{(2)}
\end{equation}
up to the second order in $E_{\rm J}/E_0$. The zeroth- and the first-order energies
are easily calculated: 
$E_{\rm CD}^{(0)} = 8E_0( \widetilde{C}_{00}^{-1} - 
\widetilde{C}_{\hat{\bf x},0}^{-1} )$
and  $E_{\rm CD}^{(1)} = 0$. To calculate the second-order term, we
apply the second-order degenerate perturbation theory, and diagonalize 
the matrix $M$, the element of which is given by 
\begin{equation} \label{eq_Mmn}
\langle i,j | M | k, l \rangle \equiv 
\sum_{ {\bf q}  \notin D} \frac{ \langle i, j | V | {\bf q} \rangle
\langle {\bf q} | V | k , l \rangle }{E_{\rm CD}^{(0)} - E_{\bf q}^{(0)} }.
\end{equation}
Here $|k,l \rangle$ is the CD states with the 
positive charge at site $k$ 
and the negative charge at site $l$, where $l$ is one of the four 
nearest neighbors of $k$,  and the sum is performed over the 
intervening states $|{\bf q} \rangle $ outside the space $D$ 
spanned by all CD states.
The above matrix element does not vanish only when $|k, l\rangle$ can be
connected to $|i,j \rangle$ by two successive charge hoppings. 
Figure~\ref{fig_cd_schem} shows all possible states of $| k, l \rangle$ 
when $|i,j \rangle$ is given as in Fig.~\ref{fig_cd_schem}(a).
While the matrix element corresponding to
Fig.~\ref{fig_cd_schem}(a) is given by
\end{multicols}
\noindent\rule{0.5\textwidth}{0.1ex}\rule{0.1ex}{2ex}\hfill
\begin{equation} \label{eq_Ma}
\langle i,j | M | i, j \rangle 
 = \frac{E_{\rm J}^2}{32 E_0} {\sum_{(k,l)}}^\prime
\frac{1}{ (\widetilde{C}^{-1}_{li} - \widetilde{C}^{-1}_{ki})
 - (\widetilde{C}^{-1}_{lj} - \widetilde{C}^{-1}_{kj})
 - (\widetilde{C}^{-1}_{00} - \widetilde{C}^{-1}_{\hat{\bf x}, 0})} ,
\end{equation}
there is only one intervening state in the case of 
Fig.~\ref{fig_cd_schem}(b), leading to
\begin{equation}
\langle i,j | M | j,i \rangle =  \frac{E_{\rm J}^2}{32 E_0}
\frac{ e^{2iA_{ji}} }
 {\widetilde{C}^{-1}_{00} - \widetilde{C}^{-1}_{\hat{\bf x}, 0}} .
\end{equation}
For $|k,l \rangle$ given by the state in Figs.~\ref{fig_cd_schem}(c) 
and (d), we find 
\begin{eqnarray}
\langle i, j|M|i, l{=}i_\alpha \rangle = \frac{E_{\rm J}^2}{32 E_0}
& & \left[   \frac{ e^{ i (A_{j,i} + A_{i,l}) }}
      {\widetilde{C}^{-1}_{00} - \widetilde{C}^{-1}_{\hat{\bf x}, 0}} +
     \frac{ e^{ i (A_{j, j_\alpha} + A_{j_\alpha,l}) }}
      {\widetilde{C}^{-1}_{\hat{\bf x} + \hat{\bf y}, 0} 
       - \widetilde{C}^{-1}_{\hat{\bf x}, 0}} +
     \frac{ e^{ i (A_{j_\alpha,l} + A_{j,j_\alpha}) }}
      {3\widetilde{C}^{-1}_{\hat{\bf x}, 0} 
       -2\widetilde{C}^{-1}_{\hat{\bf x} + \hat{\bf y}, 0} 
       -\widetilde{C}^{-1}_{00}}  \right. \nonumber \\
    &  &  + \left. \frac{ e^{i (A_{i,l} + A_{j,i}) }}
      {3\widetilde{C}^{-1}_{\hat{\bf x}, 0} 
       -\widetilde{C}^{-1}_{\hat{\bf x} + \hat{\bf y}, 0} 
       -2\widetilde{C}^{-1}_{00}} \right],
\end{eqnarray}
where $i_\alpha$  denotes the $\alpha$th
nearest neighboring site of $i$ ($\alpha = 1,2,3,4$).
Similarly, we get
\begin{eqnarray}
\langle i,j | M | k {=} j_\alpha ,j \rangle = \frac{E_{\rm J}^2}{32 E_0}
& & \left[   \frac{ e^{ i (A_{j,i} + A_{k,j}) }}
      {\widetilde{C}^{-1}_{00} - \widetilde{C}^{-1}_{\hat{\bf x}, 0}} +
     \frac{ e^{ i (A_{i_\alpha, i} + A_{k,i_\alpha}) }}
      {\widetilde{C}^{-1}_{\hat{\bf x} + \hat{\bf y}, 0} 
       - \widetilde{C}^{-1}_{\hat{\bf x}, 0}} +
     \frac{ e^{ i (A_{k, i_\alpha} + A_{i_\alpha,i}) }}
      {3\widetilde{C}^{-1}_{\hat{\bf x}, 0} 
       -2\widetilde{C}^{-1}_{\hat{\bf x} + \hat{\bf y}, 0} 
       -\widetilde{C}^{-1}_{00}}  \right. \nonumber \\
    &  &  + \left. \frac{ e^{i (A_{k,j} + A_{j,i}) }}
      {3\widetilde{C}^{-1}_{\hat{\bf x}, 0} 
       -\widetilde{C}^{-1}_{\hat{\bf x} + \hat{\bf y}, 0} 
       -2\widetilde{C}^{-1}_{00}} \right]
\end{eqnarray}
for Figs.~\ref{fig_cd_schem}(e) and (f),
\begin{equation}
\langle i,j | M | k ,l \rangle = \frac{E_{\rm J}^2}{32 E_0}
  \left[   \frac{ e^{ i (A_{j,i} + A_{k,l}) }}
      {\widetilde{C}^{-1}_{00} - \widetilde{C}^{-1}_{\hat{\bf x}, 0}} +
     \frac{ e^{ 2i (A_{k, i} + A_{j,l}) } }
      {\widetilde{C}^{-1}_{\hat{\bf x} + \hat{\bf y}, 0} 
       - \widetilde{C}^{-1}_{\hat{\bf x}, 0}} 
      + \frac{ e^{i (A_{k,l} + A_{j,i}) }}
       {2\widetilde{C}^{-1}_{\hat{\bf x} + \hat{\bf y}, 0} 
      -3\widetilde{C}^{-1}_{\hat{\bf x}, 0} 
       -\widetilde{C}^{-1}_{00}} \right]
\end{equation}
for Figs.~\ref{fig_cd_schem}(g) and (h), and finally
\begin{equation} \label{eq_Mh}
\langle i,j | M | k ,l \rangle = \frac{E_{\rm J}^2 }
{32 E_0}
\left[   \frac{ e^{ i (A_{j,i} + A_{k,l}) } }
      {\widetilde{C}^{-1}_{00} - \widetilde{C}^{-1}_{\hat{\bf x}, 0}} +
      \frac{e^{ i (A_{j,i} + A_{k,l}) }}
   { (\widetilde{C}^{-1}_{li} - \widetilde{C}^{-1}_{ki})
 - (\widetilde{C}^{-1}_{lj} - \widetilde{C}^{-1}_{kj})
 - (\widetilde{C}^{-1}_{00} - \widetilde{C}^{-1}_{\hat{\bf x}, 0})} \right] 
\end{equation}
\hfill\raisebox{-1.9ex}{\rule{0.1ex}{2ex}}\rule{0.5\textwidth}{0.1ex}
\begin{multicols}{2}
\noindent
for the cases corresponding to Fig.~\ref{fig_cd_schem}(i).
Equations~(\ref{eq_Ma})--(\ref{eq_Mh})
give the $4N{\times} 4N$ matrix $M$, which, 
again via the Fourier transformation, reduces to  
$4q^2{\times}4q^2$ Hermitian matrices for $f = p/q$. 
By diagonalizing numerically the resulting matrices, 
we obtain the second-order energy $E_{\rm CD}^{(2)}$, the
comparison of which with the energy of the 
Mott insulating phase given by Eq.~(\ref{eq_EM}) yields the
phase boundaries.  Figure~\ref{fig_cd} displays the obtained boundaries 
in the plane of $E_{\rm J}/E_1$ and $f$.
As in obtaining Fig.~3, 
we have considered systems of such large sizes that
at least three significant digits of the critical values
$(E_{\rm J}/E_1)_{\rm c}$ are obtained
for given values of $f$ and $C_1/C_0$.

\begin{figure}
\centerline{\epsfig{file=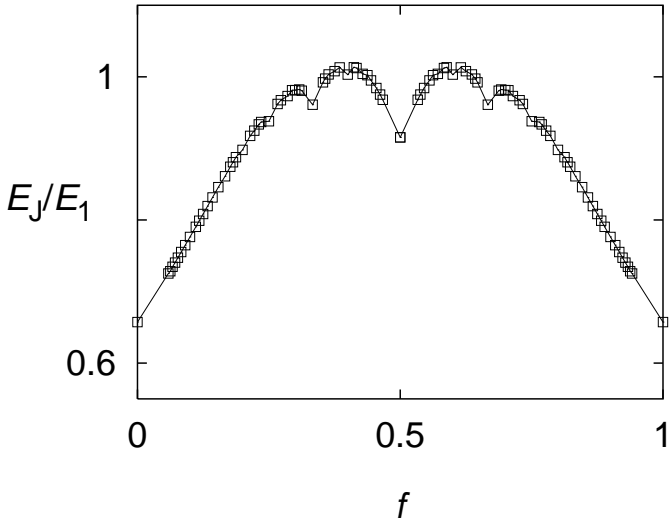,width=10.0cm}}
\vskip 0.5cm
\caption{Phase boundaries in the plane of $E_{\rm J}/E_1$ and $f$
for $C_1/C_0 = 100$ between the Mott insulating phase (below the curve)
and the superconducting phase (above) computed from the consideration 
of single-charge excitations.
The obtained critical values 
$(E_{\protect\rm J}/E_1)_{\protect\rm c} \approx 0.657$ for $f=0$ and
0.915 for $f=1/2$ are in good agreement with the experimental results.
}
\label{fig_sc1}
\end{figure}

In experiment on the arrays with $C_1/C_0 \approx 100$,
the critical values $(E_{\rm J}/E_1)_{\rm c} \approx 0.6$ for $f=0$ 
and $0.9$ for $f=1/2$ have been observed~\cite{zant};
this is to be compared with the corresponding values obtained 
in the perturbation scheme here, with the CD excitations taken into
consideration:
$(E_{\rm J}/E_1)_{\rm c} \approx 0.503$ for $f=0$ 
and $0.508$ for $f=1/2$.
Remarkably, the consideration of CD excitations yields
the critical value for $f=1/2$ not far larger than that for $f=0$,
in disagreement with the experimental result.
On the other hand, the phase boundary computed from 
the consideration of the SC excitations for
$C_1/C_0 = 100$, shown in Fig.~\ref{fig_sc1},
is in general consistent with that measured in experiment~\cite{zant}.
Contrary to the usual anticipation that
CD excitations play a crucial role in 
destroying the Mott insulating phase for $C_1\gg C_0$,
this apparently suggests that the superconductor-insulator transition 
observed in experiment is driven by SC excitations 
rather than by CD ones, raising the interesting possibility that 
the transition may not be of the BKT type.

\section{Dual Transformation Approach}

In this section, we examine the nature of the phase transitions discussed
above in terms of the topological excitations.  This is achieved by means
of the mapping of the quantum phase model given by Eq.~(\ref{eqH}) 
into an effective
3D classical model; this approach was already adopted by other authors in
the absence of the external magnetic field~\cite{fazio}.  We begin with the
Euclidean action, corresponding to the Hamiltonian in Eq.~(\ref{eqH}), in the
imaginary-time path-integral representation:
\begin{eqnarray}
S
& = & +i\sum_{j,\tau} q_{j,\tau}\nabla_\tau\phi_{j,\tau}
  + \frac{1}{2K}\sum_{ij,\tau}
    q_{i,\tau}C^{-1}_{ij}q_{j,\tau}
  \nonumber \\
& & \mbox{}
  - K\sum_{j,\mu,\tau}\cos\left[
    \nabla_\mu\phi_{j,\tau}
    -2\pi A_{\mu;j}
  \right],
  \label{QPM:S}
\end{eqnarray}
where $K\equiv\sqrt{E_{\rm J}/8E_0}$,
$A_{\mu;j}\equiv{}A_{j,j{+}\hat{\mu}}$,
$\nabla_\mu$ $(\mu=x,y)$ and
$\nabla_\tau$ are difference operators in the space and time
directions, respectively, and the (imaginary) time-slice spacing
$\delta\tau$ has been chosen to be
$\sqrt{8E_{\rm J}E_0}/\hbar$~\cite{end_note:1}.  
In the nearest-neighbor charging limit ($C_0=0$), 
the coupling constant and the time-slice spacing in Eq.~(\ref{QPM:S})
are given by $K\equiv\sqrt{E_{\rm J}/8E_1}$ and
$\delta\tau=\sqrt{8E_{\rm J}E_1}/\hbar$, respectively.
Standard procedures~\cite{Savitx80,mschoi97} then lead to
the dual form of Eq.~(\ref{QPM:S}), which is simply
 the effective Hamiltonian for the
3D classical system of vortex loops:
\end{multicols}
\noindent\rule{0.5\textwidth}{0.1ex}\rule{0.1ex}{2ex}\hfill
\begin{eqnarray}
H_V
& = & -\pi K\sum_{i,j,\tau,\tau'}\sum_{\mu=x,y,\tau}
  \left[v_\mu({\bf r}_i,\tau)-\delta_{\mu,\tau}f\right]
  \widehat {U}_\mu({\bf r}_i-{\bf r}_j,\tau-\tau')
  \left[v_\mu({\bf r}_j,\tau')-\delta_{\mu,\tau}f\right].
  \label{3DVL:H}
\end{eqnarray}
Here the vortex interaction
$
\widehat {U}_\mu({\bf r},\tau)
\equiv 2\pi\left[U_\mu(0,0)-U_\mu({\bf r},\tau)\right]
$
is determined by the Fourier transforms
\begin{eqnarray}
\widetilde{U}_x({\bf q},\omega) &=&
\widetilde{U}_y({\bf q},\omega)
 =  \frac{\widetilde{C}({\bf q})}
  {\Delta(q_x)+\Delta(q_y)+\widetilde{C}({\bf q})\Delta(\omega)}
  \\
\widetilde{U}_\tau({\bf q},\omega)
& = & \frac{1}
  {\Delta(q_x)+\Delta(q_y)+\widetilde{C}({\bf q})\Delta(\omega)}
\end{eqnarray}
\hfill\raisebox{-1.9ex}{\rule{0.1ex}{2ex}}\rule{0.5\textwidth}{0.1ex}
\begin{multicols}{2}
\noindent
with $\Delta(q)\equiv2(1-\cos{q})$.
The Fourier transform of the capacitance matrix is given by
$\widetilde{C}({\bf q})=1+(C_1/C_0)[\Delta(q_x)+\Delta(q_y)]$ 
for $C_0\neq0$ and
$\widetilde{C}({\bf q})=\Delta(q_x)+\Delta(q_y)$ for $C_0=0$.  
Note also that the vortex lines can terminate nowhere but
form closed loops or go to infinity, as implied by the condition
$\nabla\cdot{\bf{}v}({\bf r},\tau)=0$.  

The behavior of the interaction $\widehat {U}_\mu ({\bf r},\tau)$ in
Eq.~(\ref{3DVL:H}) depends crucially on whether $C_0$ vanishes, since the
Coulomb interaction between charges (Cooper pairs) is infinitely
long-ranged for $C_0=0$.  If $C_0\neq0$, one can show, in the same manner
as in Ref.~\cite{mschoi97}, that at large scales ($\sqrt{r^2+\tau^2}\gg1$)
the interaction $\widehat {U}_\mu$ is isotropic and displays the asymptotic
behavior $\widehat {U}_\mu({\bf r},\tau)\sim{-}1/\sqrt{r^2+\tau^2}$ apart from an
additive constant, regardless of the ratio $C_1/C_0$.  Accordingly, the
system is described by the 3D isotropic $XY$ model under an external
magnetic field in the $\tau$-direction, the topological representation of
which is given by Eq.~(\ref{3DVL:H}).  The 3D $XY$ model has been widely
used as a model for the bulk superconductor at temperatures low enough to
neglect the amplitude fluctuations of the order parameter~\cite{Blatte94}.
By analogy with the vortex lattice melting transition at the temperature
$T_m(H)$ in the mixed state of a type-II superconductor, a first-order
phase transition is expected at $K_c(f)$ in our system under the magnetic
field, as $K$ is increased from zero~\cite{Blatte94,end_note:2}.  At zero
field, in particular, the phase transition should be continuous, belonging
to the 3D $XY$ universality class~\cite{fazio}.

For $C_0=0$, on the other hand, the interaction is highly anisotropic:
$\widehat {U}_x({\bf r},\tau)=\widehat {U}_y({\bf r},\tau)\sim{}\exp(-\sqrt{r^2+\tau^2})$ while
$\widehat {U}_\tau({\bf r},\tau)\sim{}e^{-|\tau|}\log{r}$.  Thus the equivalent
classical system described by Eq.~(\ref{3DVL:H}) forms a layered structure
of planar spins with strongly anisotropic coupling constants.  The 3D
anisotropic $XY$ model has also been studied extensively as a special case
of the Lawrence-Doniach model for high-temperature
superconductors~\cite{Blatte94,Korshu90}: The effectively low
dimensionality enhances the phase fluctuations and lowers the transition
point $K_c(f)$~\cite{Blatte94}.  Furthermore, at zero field, the strong
anisotropy drives the transition to be of the BKT
type~\cite{fazio,Korshu90}.

\begin{figure}
\begin{center}
\centerline{\epsfig{file=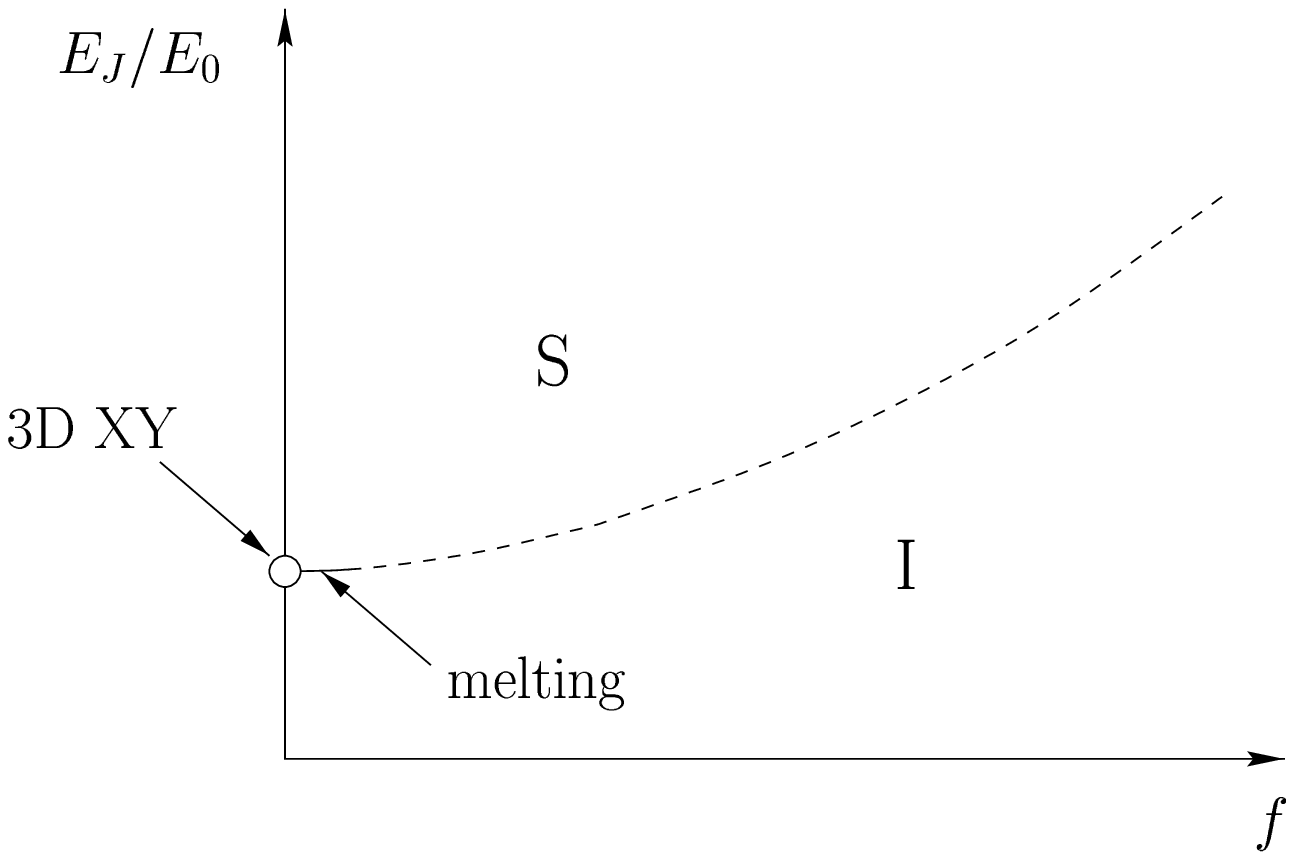,clip=,width=0.5\textwidth}}
(a)\\
\centerline{\epsfig{file=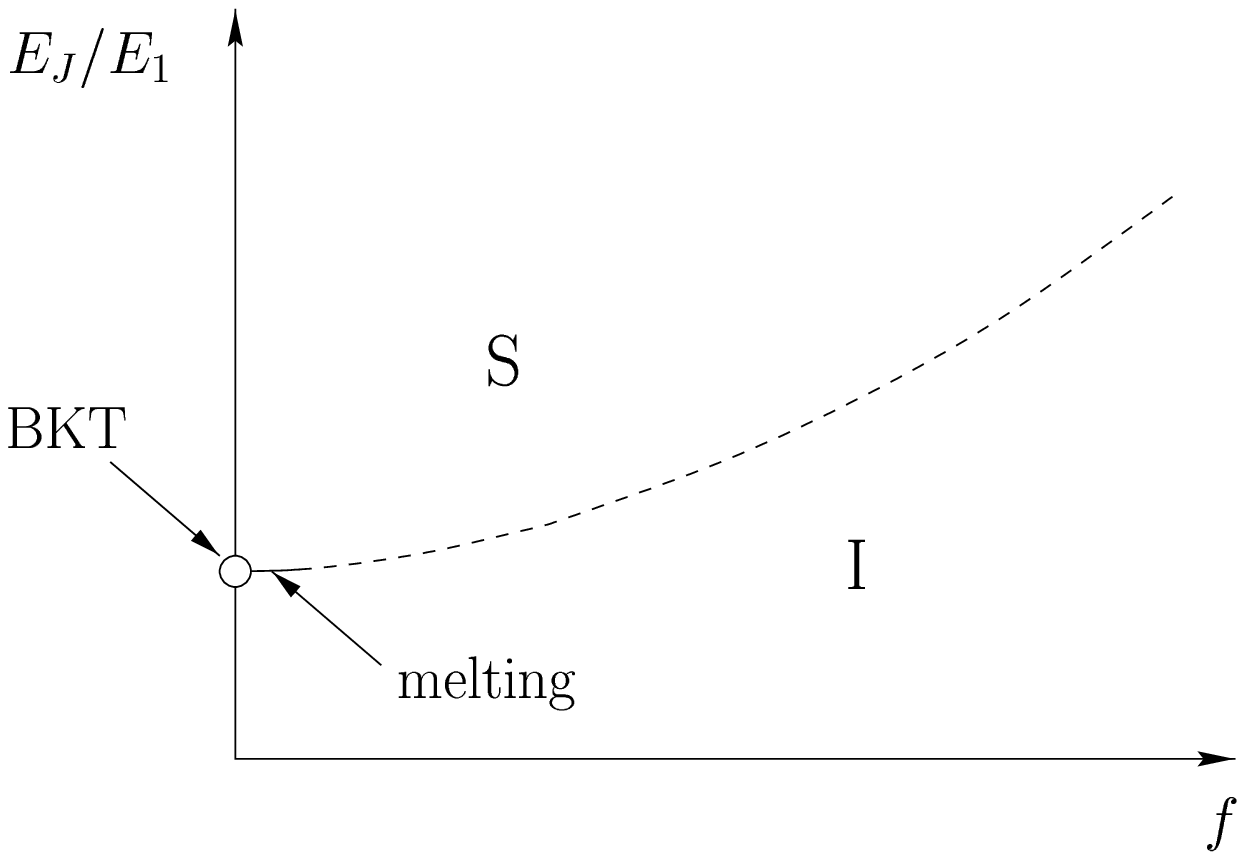,clip=,width=0.5\textwidth}}
(b)
\end{center}
\caption{Schematic diagram of the phase boundaries (a) for $C_0\neq0$
and (b) for $C_0=0$.  The dashed lines are valid only for continuum
systems.}
\label{fig:ph-dgrm}
\end{figure}

These arguments have been summarized in Fig.~\ref{fig:ph-dgrm}.  Note that
the important effects of frustration arising from the commensurability
between the flux lattice and the underlying lattice are disregarded here.
Such continuum approximation is believed to be qualitatively valid in low
field regions (represented by the solid lines in Fig.~\ref{fig:ph-dgrm}).
As the field is increased, frustration effects are expected to come into
play and to yield sensitive dependence of the transition on the field,
reproducing the perturbative results shown in Figs.~\ref{fig_sc} and
\ref{fig_cd}. 

\section{Conclusion}

We have studied the zero-temperature
phase transitions of two-dimensional superconducting arrays with both
self- and junction capacitances in the presence of external magnetic
fields.
Through the use of the second-order perturbation theory,
we have considered both single-charge excitations
and charge-dipole excitations, from which the phase diagrams are
obtained. 
It has been found that the phase boundaries are quite sensitive to 
the variation of the magnetic frustration 
due to the commensurate-incommensurate effects.
In particular, the superconductor-insulator transition 
observed in experiment has been suggested to be driven
by single-charge excitations rather than by charge-dipole ones,
and thus the possibility that the transition may not
be of the Berezinskii-Kosterlitz-Thouless type has been pointed out.
In this regard, it is noteworthy that the lowest excitation has 
already been shown 
to be comprised by the single-charge type rather than the charge-dipole type
even for large values of $C_1/C_0$, so long as
there exists finite charge frustration~\cite{b0}. 
In experiment, it is practically impossible to 
set the charge frustration exactly zero, and accordingly, 
the lowest excitation should presumably be of the single-charge type 
even in the nearest-neighbor charging limit.
Indeed it has recently been pointed out that the absence of the
Berezinskii-Kosterlitz-Thouless charge-unbinding transition
in experiments~\cite{zant} may be attributed to the presence of the
finite charge frustration which is randomly distributed over the
arrays~\cite{Delsing}.

We have also transformed the
system to a 3D classical $XY$ model under magnetic fields in the
time direction. 
The nature of the phase transitions at low magnetic fields
has been discussed, based on the analogy with the bulk superconductor under
magnetic fields.
Unfortunately, unlike in the 2D $XY$ model, 
there have been few studies of the
frustration effects in the 3D $XY$ model, 
which disallows quantitative comparison at this stage.
Nevertheless, the analogy with the continuum superconductor
provides a complement to the perturbative estimate of the phase
boundaries, already giving a clue to the nature of the phase transitions.

\acknowledgements
This work was supported in part by the Basic Science Research Institute 
Program, Ministry of Education, in part by the Korea Science and Engineering
Foundation through the SRC Program.  MSC was also supported in part by the 
Ministry of Science and Technology
through the Creative Research Initiative Program.

\narrowtext
\vskip 1cm

\vskip 1cm

\vskip 5cm

\vskip 4cm

\end{multicols}

\begin{references}
\bibitem{end_note:2}
The upper critical field $H_{c2}$ for a type-II superconductor 
is irrelevant in our consideration here; ideally, the system 
corresponds to the limit $H_{c2}\to\infty$.
\end{references}
\end{document}